\documentclass[11pt]{article}

\usepackage[preprint]{acl}

\usepackage{times}
\usepackage{latexsym}

\usepackage[T1]{fontenc}

\usepackage[utf8]{inputenc}

\usepackage{microtype}

\usepackage{inconsolata}

\usepackage{graphicx}
\usepackage{amsmath}
\usepackage{mathtools}
\usepackage{amsthm}
\usepackage{booktabs} 
\usepackage[most]{tcolorbox}
\usepackage{listings}
\usepackage{xcolor}
\usepackage{multirow}
\usepackage[table]{xcolor}
\usepackage{geometry} 
\usepackage{threeparttable}
\usepackage{pifont}
\newcommand{\cmark}{\ding{51}} 
\newcommand{\xmark}{\ding{55}} 
\usepackage{tabularx}
\usepackage{array}
\usepackage{adjustbox} 

\usepackage{float,comment}

\usepackage{hyperref}
\usepackage{amsfonts}
\usepackage{algorithm}
\usepackage{algorithmicx}
\usepackage{algpseudocode}
\usepackage[caption=false,font=normalsize,labelfont=sf,textfont=sf]{subfig}
\usepackage{textcomp}
\usepackage{stfloats}
\usepackage{url}
\usepackage{xcolor}
\usepackage{verbatim}
\usepackage{makecell}
\usepackage{bbding}
\usepackage{changepage}

\lstdefinelanguage{XMLStyle}{
  basicstyle=\ttfamily\footnotesize,
  morestring=[b]",
  morestring=[s]{>}{<},
  morecomment=[s]{<?}{?>},
  stringstyle=\color{black},
  identifierstyle=\color{blue}, 
  keywordstyle=\color{cyan},
  breaklines=true,
  frame=single,   
  backgroundcolor=\color{gray!5}, 
}

\title{DM-ASR: Diarization-aware Multi-speaker ASR with Unified Offline and Online Decoding with Large Language Models}

\author{
  \textbf{Li Li}\textsuperscript{1,2},
  \textbf{Ming Cheng}\textsuperscript{3},
  \textbf{Weixin Zhu}\textsuperscript{4},
  \textbf{Yannan Wang}\textsuperscript{4},
  \textbf{Juan Liu}\textsuperscript{1},
  \textbf{Ming Li}\textsuperscript{1,2}\thanks{Corresponding author.}
  \\[5pt]
  \textsuperscript{1}School of Artificial Intelligence, Wuhan University
  \\
  \textsuperscript{2}School of Artificial Intelligence, The Chinese University of Hong Kong, Shenzhen
  \\
  \textsuperscript{3}School of Computer Science, Wuhan University
  \\
  \textsuperscript{4}Tencent Ethereal Audio Lab, Tencent
  \\
  \small
  \href{mailto:mingli369@cuhk.edu.cn}{mingli369@cuhk.edu.cn}
}
  
\begin{document}
\maketitle
\begin{abstract}
Multi-speaker automatic speech recognition (ASR) aims to recognize not only what was said, but also who said it and when. Recent Speech-LLMs have shown promising results in structured multi-speaker transcription, but jointly modeling speaker attribution, temporal information, and lexical content remains challenging and often requires substantial model capacity and training data. To address this challenge, we propose DM-ASR, a \textbf{D}iarization-aware \textbf{M}ulti-speaker \textbf{ASR} framework that reformulates multi-speaker recognition as diarization-conditioned multi-turn dialogue generation. By leveraging diarization-derived speaker-time cues as explicit structural priors, DM-ASR decomposes transcription into a sequence of conditioned queries while preserving mixed acoustic information and conversational context. We further introduce optional word-level timestamps, fine-tune the model with Group Sequence Policy Optimization (GSPO), and analyze the complementary roles of diarization cues and language-model predictions. The proposed formulation also enables unified offline and causal segment-level online decoding through flexible acoustic context and dialogue history management. Experiments on Mandarin and English benchmarks demonstrate competitive structured transcription performance with relatively compact models and training data, while effectively supporting both offline and online recognition.
\end{abstract}

\section{Introduction}
Multi-speaker ASR aims to recover what was said, by whom, and when in conversations such as meetings, interviews, and telephone calls~\cite{alimeeting,aishell4,ami}. Unlike single-speaker ASR, it must jointly handle lexical recognition, speaker attribution, and temporal grounding amid turn-taking, overlapping speech, and long-range conversational context.

Existing systems for multi-speaker ASR can be broadly grouped into four categories. The first is \emph{cascaded diarization--ASR}, where diarization estimates speaker turns before ASR transcribes the resulting segments~\cite{sd+asr2_2021,sd+asr1_2024,sd_asr3_2022,sd_asr4_2024}. These modular pipelines are practical, but diarization errors can propagate to the transcript, and hard segmentation can lose acoustic or conversational context~\cite{sd4_2025,sd_asr5_2023}. The second category is \emph{jointly trained speaker-attributed ASR}, which more tightly couples speaker and lexical prediction to reduce the mismatch between separate modules~\cite{wzhang_multispeaker,sot,lstm-transformer,alimeeting_2023,parksortformer,wang2025meta,zheng2025dncasr}. The third is \emph{LLM-assisted hybrid systems}: these retain a multi-component pipeline and use an LLM to refine diarization, speaker attribution, or transcript consistency~\cite{wang2024diarizationlm,sd_llm_3_2024,sd_llm_2_2025}. The fourth is \emph{unified end-to-end Speech-LLMs}, which formulate multi-speaker transcription as structured generation from speech~\cite{10887642,yin2025speakerlm,shi2025train,huo2026tagspeech,yu2026moss,peng2026vibevoiceasrtechnicalreport,tu2026vibevoiceasrstreaming}.

\newcolumntype{L}{>{\raggedright\arraybackslash}X}
\newcolumntype{C}{>{\centering\arraybackslash}p{2.1cm}}
\newcolumntype{O}{>{\raggedright\arraybackslash}p{1.15cm}}
\begin{table*}[t]
\centering
\small
\setlength{\tabcolsep}{1.5pt}
\renewcommand{\arraystretch}{1.2}
\caption{Comparison of recent works on multi-speaker ASR by explicit text outputs and whether support online ASR.}
\label{tab:task_comparison}
\begin{tabularx}{\textwidth}{L C C C C O}
\toprule
\textbf{Recent Work on Multi-speaker ASR} &
\textbf{Use LLM} &
\textbf{Speaker} &
\textbf{Segment-level Timestamp} &
\makecell[tl]{\textbf{Word-level}\\\textbf{Timestamp}} &
\makecell[tl]{\textbf{Online}\\\textbf{ASR}}\\
\midrule
Sortformer \cite{parksortformer}; Meta-CAT \citep{wang2025meta}; DNCASR~\citep{zheng2025dncasr}
& \xmark & \cmark & \xmark & \xmark & \xmark\\
MT-LLM \citep{meng2025large}; DiarizationLM \citep{wang2024diarizationlm} {\scriptsize (transcript reconciliation)}
& \cmark & \xmark & \xmark & \xmark & \xmark\\
SpeakerLM \citep{yin2025speakerlm}
& \cmark & \cmark & \xmark & \xmark & \xmark\\
JEDIS-LLM \citep{shi2025train}
& \cmark & \cmark & \xmark & \xmark & \cmark\\
TagSpeech \cite{huo2026tagspeech}; VibeVoice-ASR \cite{peng2026vibevoiceasrtechnicalreport}; MOSS Transcribe Diarize \cite{yu2026moss}
& \cmark & \cmark & \cmark & \xmark & \xmark\\
VibeVoice-ASR-Streaming \cite{tu2026vibevoiceasrstreaming}
& \cmark & \cmark & \xmark & \xmark & \cmark\\
\midrule
\rowcolor{gray!15}
\textbf{DM-ASR (Ours)} & \textbf{\cmark} & \textbf{\cmark} & \textbf{\cmark} & \textbf{\cmark} & \textbf{\cmark} \\
\bottomrule
\end{tabularx}
\end{table*}

Despite this progress, temporal output remains unevenly covered: many systems focus on \emph{who said what}, while recent Speech-LLMs have begun to generate segment-level timestamps and word-level timing remains less explored~\cite{parksortformer,wang2025meta,zheng2025dncasr,huo2026tagspeech,yu2026moss,peng2026vibevoiceasrtechnicalreport,whisper}. Unified Speech-LLMs are promising, but learning lexical, speaker, and temporal structure together can require substantial model capacity and training data. Moreover, online use imposes causal constraints: the system must decode without future audio, and strong end-to-end streaming performance remains an open challenge even as streaming-oriented models emerge~\cite{shi2025train,tu2026vibevoiceasrstreaming}. These considerations make explicit diarization cues a useful structural prior, particularly when data or model capacity is limited.

Motivated by these observations, we propose \textbf{DM-ASR}. DM-ASR reformulates multi-speaker speech recognition as diarization-conditioned multi-turn generation, enabling structured, context-aware transcription under a unified offline and online inference framework. Unlike a conventional cascade, it does not cut the recording into speaker-isolated waveforms. Instead, external diarization supplies speaker--time queries to a speech encoder and language-model decoder operating on shared mixed audio, allowing the model to recognize each target while retaining conversational context from preceding turns. Optional word--timestamp targets provide finer temporal supervision, and perturbations to input speaker labels and boundaries expose the model to imperfect diarization cues. We also fine-tune the model with Group Sequence Policy Optimization (GSPO) to improve speaker-attributed transcription. To extend this formulation to online use, we support segment-level causal decoding. Once the diarization front end finalizes a target segment, the model transcribes it using mixed-audio context that ends at the segment endpoint, together with preceding recognized turns as dialogue context. This lets the model commit each response without access to future audio while preserving conversational history.

Our contributions are threefold. (1) \textbf{Diarization-conditioned multi-turn formulation:} we decompose multi-speaker ASR into speaker--time-conditioned generation turns, preserving the shared mixed audio and cross-turn context. (2) \textbf{Structured and robust transcription:} we incorporate word-level timestamp supervision and sequence-level GSPO optimization, and analyze speaker/time cues to study model behavior under imperfect diarization.
(3) \textbf{Unified offline and online inference:} the same trained model and generation mechanism support offline and causal segment-level online decoding.

\section{Related Work}
\subsection{Speaker Diarization Systems}

Speaker diarization aims to determine \emph{who spoke when} in multi-speaker speech and serves as a fundamental component of conversational speech understanding \cite{kanda2020joint,park2022review}. Early studies on speaker diarization focus on modularized pipelines~\cite{wang2018speaker,lin2019lstm,landini2022bayesian}, which first segment the audio and then cluster the segments based on speaker embedding similarity. Since these methods typically assume that each segment contains only one active speaker, their performance degrades substantially in overlapping speech. To address this limitation, End-to-End Neural Diarization (EEND) formulates diarization as a multi-label prediction problem, enabling more robust modeling of overlapping speech~\cite{fujita2019end_1,horiguchi2020end,horiguchi2022encoder}. Building on this line of work, Pyannote~\cite{pyannote_1_2020,pyannote_2_2023} has shown strong performance across multiple benchmarks. DiariZen~\cite{diarizen_1_2025,diarizen_2_2025} further improves upon Pyannote by incorporating pretrained WavLM representations~\cite{wavlm_2022}, thereby benefiting from richer acoustic information. In parallel, Target-Speaker Voice Activity Detection (TSVAD) combines modular and neural designs and has achieved strong performance in practical diarization systems~\cite{medennikov2020target,cheng2023dku,wang2022similarity,wang2021dku}. More recently, Sequence-to-Sequence Neural Diarization (S2SND)~\cite{cheng2024sequence,Cheng2025MultiChannelSN} has advanced the online diarization setting through a sequence generation formulation.

\subsection{Multi-speaker ASR: Cascaded and Joint Modeling}

A common solution to multi-speaker ASR is the cascaded pipeline~\cite{sd+asr2_2021,sd+asr1_2024,sd_asr3_2022,sd_asr4_2024}, in which speaker diarization is first performed to estimate speaker turns and boundaries, and ASR is then applied to the resulting segments. While effective and easy to deploy, such systems suffer from error propagation, speaker-text mismatch, and limited ability to exploit the mutual dependence between speaker structure and lexical content ~\cite{sd4_2025,sd_asr5_2023}. These issues become more severe in overlapping, noisy, or long-form conversations.

To mitigate these limitations, a line of work has moved toward jointly trained speaker-attributed ASR, where speaker modeling and recognition are optimized in a more unified manner. Early efforts explored joint speech recognition and speaker diarization through sequence transduction and related integrated formulations~\cite{kanda2020joint}. This direction was later extended to end-to-end and transcribe-to-diarize modeling, further tightening the coupling between speaker modeling and recognition~\cite{sot,lstm-transformer,alimeeting_2023,parksortformer,wang2025meta,zheng2025dncasr}. As summarized in Table~\ref{tab:task_comparison}, systems such as Sortformer~\cite{parksortformer}, Meta-CAT~\cite{wang2025meta}, and DNCASR~\cite{zheng2025dncasr} primarily target \emph{who said what}, rather than the fuller \emph{who said what when} objective.

\subsection{LLM-based Multi-speaker ASR}

Recent advances in large language models (LLMs) have opened a new direction for multi-speaker ASR. One line uses LLMs to refine or reconcile outputs from conventional diarization--ASR pipelines. For example, DiarizationLM improves transcript consistency and speaker attribution by refining the outputs of independent ASR and diarization systems~\cite{wang2024diarizationlm}. As shown in Table~\ref{tab:task_comparison}, these LLM-assisted pipelines can leverage language-based reasoning, but remain dependent on upstream components and do not explicitly generate speaker-aware timestamps.

\paragraph{Unified offline Speech-LLMs.} A second line formulates multi-speaker ASR as structured generation directly from speech, allowing a unified model to use conversational context. Representative systems include SpeakerLM, TagSpeech, MOSS Transcribe Diarize, and VIBEVOICE-ASR~\cite{yin2025speakerlm,huo2026tagspeech,yu2026moss,peng2026vibevoiceasrtechnicalreport}. Their output scope varies: SpeakerLM primarily addresses \emph{who said what}, whereas TagSpeech, MOSS Transcribe Diarize, and VIBEVOICE-ASR also generate segment-level timestamps, moving toward \emph{who said what when}~\cite{yin2025speakerlm,huo2026tagspeech,yu2026moss,peng2026vibevoiceasrtechnicalreport}.

\paragraph{Streaming Speech-LLMs.} Complementing these offline systems, recent work studies incremental speaker-attributed transcription. JEDIS-LLM uses a Speaker Prompt Cache to maintain speaker consistency during streaming generation, but does not explicitly predict timestamps~\cite{shi2025train}. VIBEVOICE-ASR-Streaming processes 2.9\,s audio chunks with 0.5\,s lookahead and emits speaker-attributed text~\cite{tu2026vibevoiceasrstreaming}.

Despite these advances, existing Speech-LLMs largely rely on
learning speaker and temporal structure jointly with transcription,
while online variants primarily focus on incremental decoding.
How to effectively leverage explicit diarization cues within
a unified generation framework that supports both offline and
causal online recognition remains underexplored.
Motivated by this gap, DM-ASR uses diarization-derived
speaker-time queries to organize multi-speaker transcription
as multi-turn generation, combining explicit structural
guidance with contextual language modeling and flexible inference.

\begin{figure*}[t]
\centering
\IfFileExists{framework.png}{\includegraphics[width=\linewidth]{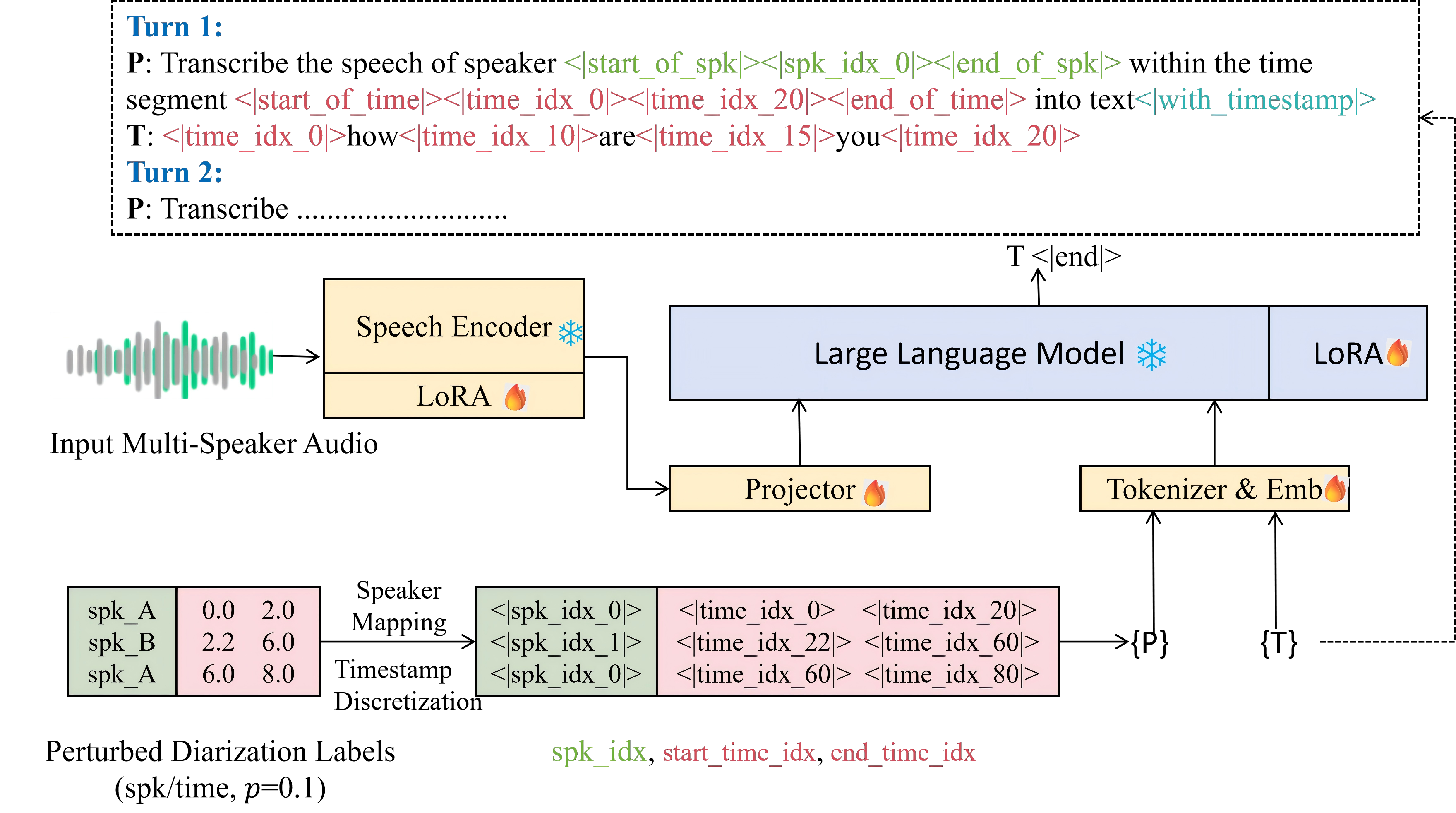}}{\fbox{\parbox{0.94\linewidth}{\centering Framework figure: add \texttt{framework.png}}}}
\caption{Overview of DM-ASR. Diarization-derived speaker--time queries guide multi-turn transcription of mixed conversational audio. Speaker labels and segment boundaries are perturbed with probability $p=0.1$ before speaker mapping and timestamp discretization.}
\label{fig:framework}
\end{figure*}

\section{Method}
\subsection{Overview and Problem Formulation}
\paragraph{Problem Formulation.}
Given mixed conversational audio $X$ and diarization output $D=\{(s_k,t_k^{\mathrm{st}},t_k^{\mathrm{ed}})\}_{k=1}^{K}$, DM-ASR predicts a speaker-attributed transcription. Here, $s_k$ is a speaker identity and $[t_k^{\mathrm{st}},t_k^{\mathrm{ed}}]$ is the $k$-th diarized interval. We use $D$ as a \emph{structural prior} to form queries, rather than cutting $X$ into isolated speaker waveforms. For query $q_k$, the model generates response $r_k$ conditioned on the shared audio representation and preceding dialogue turns $H_{<k}$:
\begin{equation}
 P_\theta(R\mid X,D)=\prod_{k=1}^{K}P_\theta(r_k\mid X,q_k,H_{<k}),
 \label{eq:multi_turn}
\end{equation}
where $R=(r_1,\ldots,r_K)$.

\paragraph{Framework Overview.}
To realize this formulation, DM-ASR uses four components, illustrated in Figure~\ref{fig:framework}: (1) a speech encoder that extracts acoustic representations from the mixed multi-speaker audio; (2) a projector that maps these representations into the LLM embedding space; (3) an autoregressive LLM decoder that generates each response conditioned on its query and preceding turns; and (4) a tokenization mechanism that converts diarization outputs into speaker and timestamp tokens to construct the queries.

\subsection{Diarization-aware Prompt Construction}
Each diarized segment is converted to a speaker--time query. Speaker labels are remapped to local indices within the acoustic window, with the mapping retained to recover recording-level identities. Timestamps are expressed relative to the start of the acoustic window and discretized as $u=\mathrm{round}(t/\Delta t)$ with $\Delta t=0.1$\,s, yielding \texttt{<|time\_idx\_0|>}--\texttt{<|time\_idx\_250|>} for a 25\,s window; four additional tokens represent 0.02--0.08\,s offsets near the window start. Speaker labels are remapped to at most 20 local indices per window. The prompt asks the model to transcribe the specified speaker within the specified interval. This provides an explicit target without removing other speakers' acoustic context from audio. The lower part of Figure~\ref{fig:framework} illustrates how diarization labels are mapped and discretized into the speaker--time tokens used to form these queries. Appendix~\ref{app:generation} lists the token inventory and prompt templates.

\subsection{Multi-turn Dialogue Generation}
For an acoustic window containing $K$ diarized segments, we form an ordered sequence of query--response turns $\{(q_k,r_k)\}_{k=1}^{K}$. The first turn includes the audio representation and its speaker--time query; later turns introduce a new query while retaining previous dialogue context. This makes the number of generated turns follow the diarization output and lets the decoder condition each response on earlier recognized content. The upper part of Figure~\ref{fig:framework} sketches this sequence, showing the initial audio-conditioned turn followed by a subsequent dialogue turn. The query asks the model to transcribe the target and to correct its speaker label; each response therefore begins with a speaker token, followed by the transcript, and, when \texttt{<|with\_timestamp|>} is appended to the query, by interleaved timestamp tokens. The several evaluation choices for final output fields are described in Appendix~\ref{app:cue_analysis}.

\subsection{Training and Temporal Supervision}
Training concatenates query--response turns and uses teacher forcing, with cross-entropy applied only to supervised response tokens:
\begin{equation}
\mathcal{L}_{\mathrm{gen}}=-\sum_{k=1}^{K}\sum_{j=1}^{|r_k|}\log P_\theta(r_{k,j}\mid X,q_k,H_{<k},r_{k,<j}).
\end{equation}
For word-level temporal supervision, forced-aligned word boundaries are serialized with the transcript when the \texttt{<|with\_timestamp|>} control token is used:
\begin{equation}
r_k=[\tau_1,w_1,\tau_2,w_2,\ldots,\tau_n,w_n,\tau_{n+1}],
\end{equation}
where $w_i$ and $\tau_i$ denote a lexical unit and its boundary time. For Mandarin, $w_i$ is a character; for English, a word. In the perturbed condition, the input speaker label and each interval boundary are perturbed with probability $p=0.1$ (boundaries by $\pm0.5$\,s), while response targets remain unchanged. The lower-left of Figure~\ref{fig:framework} illustrates this perturbation path from diarization labels to query construction, separating noisy conditioning from the transcription target.

\subsection{Sequence-level Optimization with GSPO}
\label{sec:gspo}
Cross-entropy training uses token-level log-likelihoods, whereas evaluation is performed on the complete speaker-attributed transcript. To reduce this mismatch, we further fine-tune DM-ASR with GSPO~\cite{zheng2025gspo}, regarding the model as a policy $\pi_\theta$ with parameters $\theta$.

\textbf{Sampling and reward.}
For each training window $x$ (up to 25\,s of audio with its turn prompts), the current model $\pi_{\theta_{\mathrm{old}}}$ samples a group of $G=8$ responses $\{y_i\}_{i=1}^{G}$ (temperature 1.0, top-$p$ 1.0). Each response is scored by the metrics used at evaluation:
\begin{equation}
r_i=-\bigl(0.6\,\mathrm{cpCER}(y_i)+0.4\,\mathrm{tcpCER}(y_i)\bigr),
\end{equation}
where tcpCER uses the predicted timestamps with a 5-s collar, and cpWER/tcpWER are used for English. Since the responses share the same input, their relative quality is measured by the group-normalized advantage
$\hat{A}_i=(r_i-\bar{r})/(\sigma_r+10^{-6})$, where $\bar{r}$ and $\sigma_r$ are the mean and standard deviation of $\{r_j\}_{j=1}^{G}$; $\hat{A}_i=0$ if all rewards are equal.

\textbf{Sequence-level objective.}
GSPO raises the likelihood of responses with positive advantage and lowers it for negative ones, measuring the policy change at the sequence level. Let $\ell_\theta(y_i)=|y_i|^{-1}\log\pi_\theta(y_i\mid x)$ be the length-normalized log-likelihood over the $|y_i|$ non-speaker tokens of $y_i$. The importance ratio $s_i=\exp\bigl(\ell_\theta(y_i)-\ell_{\theta_{\mathrm{old}}}(y_i)\bigr)$ compares the updated and sampling policies, and is clipped to $\tilde{s}_i=\mathrm{clip}(s_i,1-\varepsilon_{\mathrm{low}},1+\varepsilon_{\mathrm{high}})$ to keep each update small. The objective is
\begin{equation}
\mathcal{J}(\theta)=\mathbb{E}_{x}\Bigl[\frac{1}{G}\sum_{i=1}^{G}\min\bigl(s_i\hat{A}_i,\,\tilde{s}_i\hat{A}_i\bigr)\Bigr],
\end{equation}
with $\varepsilon_{\mathrm{low}}=3\times10^{-4}$ and $\varepsilon_{\mathrm{high}}=4\times10^{-4}$. No KL or entropy term is used.

\subsection{Unified Offline and Online Inference}
The same diarization-conditioned multi-turn generation procedure supports both offline and online inference. The inference regimes differ in the acoustic context available to each target and in when that target is processed.

\paragraph{Offline decoding.} The recording is processed in acoustic windows of at most $L=25$\,s; within each window the model generates an ordered sequence of query--response turns, using preceding recognized turns as dialogue context.

\paragraph{Online segment-level decoding.} Processing is triggered when the diarization front end finalizes target segment $k$. The target speaker follows the front-end diarization mapping, with local window indices mapped back to recording-level labels. Given endpoint $t_k^{\mathrm{ed}}$, DM-ASR processes the mixed audio in a strictly causal window:
\begin{equation}
W_k=[\max(0,t_k^{\mathrm{ed}}-L),t_k^{\mathrm{ed}}]
\end{equation}
The model generates the target transcription from this audio and only previously committed turns whose start times fall within $W_k$; older turns are removed from the dialogue history. After generation, the response is committed and appended to history for subsequent targets. Thus, the decoder uses no audio beyond the target endpoint and no uncommitted responses. Appendix~\ref{app:online_decoding}, Algorithm~\ref{alg:online_decoding}, details the finalization and history-update procedure.

\begin{table*}[t]
	\centering
	\setlength{\tabcolsep}{2pt}
	\renewcommand{\arraystretch}{0.8}
	\caption{Offline multi-speaker ASR results on AliMeeting and AISHELL-4.
Metrics are DER (\%), cpCER (\%), and tcpCER (\%). DER reported in VibeVoice-ASR and Qwen-Audio-3.1-ASR-Next uses a 0.5-s collar, while other methods, use no collar (0 s). For fair comparison, we also report 0.5-s-collar DER in parentheses. Qwen-Audio-3.1-ASR-Next is evaluated on approximately 4-minute chunks, whereas all DM-ASR results are computed on complete recordings.}
	\label{exp:man}
	\begin{threeparttable}[b]
    \resizebox{1\textwidth}{!}{
	\begin{tabular}{llrrrrrr}
		\toprule
		\multirow{3}{*}{\textbf{Method}}
		& \multirow{3}{*}{\textbf{\makecell[l]{Params}}}
		& \multicolumn{3}{c}{\textbf{AliMeeting}}& \multicolumn{3}{c}{\textbf{AISHELL-4}} \\
		\cmidrule(l{2pt}r{2pt}){3-5} \cmidrule(l{2pt}r{2pt}){6-8} 
		&&\textbf{\makecell[l]{DER (\%)}}&\textbf{\makecell[l]{cpCER (\%)}} & \textbf{\makecell[l]{tcpCER (\%)}}
        &\textbf{\makecell[l]{DER (\%)}}&\textbf{\makecell[l]{cpCER (\%)}} & \textbf{\makecell[l]{tcpCER (\%)}} \\
        \midrule
        \textbf{Cascade Baselines} \\
        \midrule
        Pyannote+Whisper-large-v3 & 1.5B  & 26.13 & 46.56 & - & - & - &- \\
        DiariZen+Whisper-large-v3 & 1.5B  & 10.80 & 41.05 & 43.75 & 11.97 & 34.42 & 37.40 \\
	    \midrule
        \textbf{End-to-End Baselines} \\
        \midrule 
        Qwen2.5-Omni-7B~\cite{huo2026tagspeech} &  7B & 37.42 & 41.23 & - & - & - & - \\
        Gemini-2.5-Pro~\cite{peng2026vibevoiceasrtechnicalreport} & - & 31.60 & 41.64 & 53.49& 15.32 & 31.59 & 35.96 \\
        Gemini-3.0-Pro~\cite{peng2026vibevoiceasrtechnicalreport} & -  & 38.75 & 32.84 & 65.61 & 22.03 & 27.43 & 54.17 \\
        Qwen-Audio-3.1-ASR-Next$^{*}$~\cite{qwen_audio_3_1_asr_report} & -- & \cellcolor{gray!25}(2.15) & \cellcolor{gray!25}12.04 & -- & \cellcolor{gray!25}(2.70) & \cellcolor{gray!25}13.51 & -- \\
        \midrule
        SpeakerLM (694h)~\cite{yin2025speakerlm} & 7B  & - & 29.60 & - & - & 25.28 & - \\
        SpeakerLM (2270h)~\cite{yin2025speakerlm} & 7B  & - & 27.97 & - & - & 23.10 & - \\
        SpeakerLM (7639h)~\cite{yin2025speakerlm} & 7B  & - & 16.05 & - & - & 18.37 & - \\
        Tagspeech(103h)~\cite{huo2026tagspeech} &  7B & 22.13 & 33.84 & - & - & - & - \\
        MOSS Transcribe Diarize $^{*}$~\cite{yu2026moss} & 0.6B & 12.50 & 15.82 & \cellcolor{gray!25}16.02 & 9.56 & 15.83 & 16.22 \\
        VibeVoice-ASR(9400+h)~\cite{peng2026vibevoiceasrtechnicalreport} & 7B  & (10.92) & 29.33 & 29.51 &(6.77) & 24.99 & 25.35 \\
        \midrule
        \multicolumn{8}{l}{\textbf{Ours} (diarization-provided speakers; LLM-predicted times)} \\
        \midrule
        {DM-ASR} (DiariZen) (CN 630h) & 0.6B & 11.00 & 23.46 & 24.09 & 12.01 & 21.58 & 23.04 \\
        {DM-ASR} (DiariZen) (CN 1300h) & 0.6B & 10.97 & 21.60 & 22.23 & 12.00 & 19.69 & 20.40 \\
        {DM-ASR} (DiariZen) (CN+EN 2900h) & 0.6B & 10.94 & 21.73 & 22.35 & 12.01 & 18.96 & 19.67 \\
        {DM-ASR} (DiariZen) (CN 1300h) & 1.7B & 10.92 & 20.95 & 21.55 & 12.01 & 19.85 & 20.79 \\
        {DM-ASR} (S2SND) (CN 1300h) & 1.7B & 10.09 & 19.15 & 19.45 & 10.56 & 19.21 & 19.94 \\
        {DM-ASR} (DiariZen) (CN+EN 2900h) & 1.7B & 10.92 & 23.32 & 24.02 & 12.01 & 19.01 & 19.73 \\
        {DM-ASR} (S2SND) (CN+EN 2900h) & 1.7B & 10.09 (2.23) & 21.40 & 21.79 & 10.56 (4.16) & 17.66 & 18.10 \\
        {DM-ASR-L} (S2SND) (CN+EN 7500h) & 4.0B & 10.08 (2.20) & 17.51 & 17.78 & 9.70 (3.39) & 15.58 & 15.95 \\
        \quad + GSPO  (S2SND) (CN+EN 7500h) & 4.0B & 10.09 (2.22) & 16.77 & 17.00 & 9.71 (3.39) & 14.74 & \cellcolor{gray!25}15.03 \\
        \bottomrule
    \end{tabular}}
    \begin{tablenotes}[flushleft]
    \footnotesize
    \item[*]\parbox[t]{0.94\textwidth}{MOSS and Qwen-Audio-3.1-ASR-Next use extensive training data, but neither reports the total training hours.}
    \end{tablenotes}
	\end{threeparttable}
\end{table*}

\begin{table*}[t]
	\centering
	\setlength{\tabcolsep}{2pt}
	\renewcommand{\arraystretch}{1.0}
	\caption{Offline multi-speaker ASR results on AMI-IHM, AMI-SDM, and Fisher.
Metrics are DER (\%), cpWER (\%), and tcpWER (\%). DER reported in VibeVoice-ASR uses a 0.5-s collar, while other methods, use no collar (0 s). For fair comparison, we also report 0.5-s-collar DER in parentheses.}
	\label{exp:eng}
	\begin{threeparttable}[b]
    \resizebox{1\textwidth}{!}{
	\begin{tabular}{llrrrrrrrrr}

		\toprule
		\multirow{3}{*}{\textbf{Method}}
		& \multirow{3}{*}{\textbf{Params}}
		& \multicolumn{3}{c}{\textbf{AMI-IHM}}& \multicolumn{3}{c}{\textbf{AMI-SDM}}& \multicolumn{3}{c}{\textbf{Fisher}} \\
		\cmidrule(l{2pt}r{2pt}){3-5} \cmidrule(l{2pt}r{2pt}){6-8} \cmidrule(l{2pt}r{2pt}){9-11}
		& & \textbf{DER (\%)} & \textbf{cpWER (\%)} & \textbf{tcpWER (\%)}
          & \textbf{DER (\%)} & \textbf{cpWER (\%)} & \textbf{tcpWER (\%)}
          & \textbf{DER (\%)} & \textbf{cpWER (\%)} & \textbf{tcpWER (\%)} \\
        \midrule
        \textbf{Cascade Baselines} \\
        \midrule
        Pyannote + Whisper-large-v3 & 1.5B & - & - & - & 23.05 & 43.57 & - & - & - & - \\
        DiariZen + Whisper-large-v3 &  1.5B & 27.67 & 32.27& 33.99& 14.61 & 43.91 & 49.06 & 24.72 & 26.29&26.75\\
	    \midrule
        \textbf{End-to-End Baselines} \\
        \midrule 
        Qwen2.5-Omni-7B~\cite{huo2026tagspeech} & 7B & - & - & - & 34.71 & 49.86 & - & - & - & - \\
        Gemini-2.5-Pro~\cite{peng2026vibevoiceasrtechnicalreport} & - &23.54 & 29.57 & 38.35& 23.79 & 34.78 & 41.39 & - & - & -\\
        Gemini-3.0-Pro~\cite{peng2026vibevoiceasrtechnicalreport} & -  & 46.23 & 22.34 & 63.65 & 43.04 & 26.91 & 64.86& - & - & - \\
        \midrule
        JEDIS-LLM (10000h)~\cite{shi2025train} & 5.6B & - & 23.13 & - & - & - & - & - & \cellcolor{gray!25}15.03 & - \\
        TagSpeech (65h)~\cite{huo2026tagspeech} & 7B & - & - & - & 24.84 & 42.55 & - & - & - & - \\
        VibeVoice-ASR~\cite{peng2026vibevoiceasrtechnicalreport} & 7B  & (11.92) & 20.41 & 20.82 &(13.43) & 28.82 & 29.80& - & - & - \\
        \midrule
        \multicolumn{11}{l}{\textbf{Ours} (diarization-provided speakers; LLM-predicted times)} \\
        \midrule
        {DM-ASR} (DiariZen) (EN 630h) & 0.6B & 27.87 & 18.45 & 19.52 & 14.83 & 26.89 & 28.90 & 24.71 & 17.29 & 17.43 \\
        {DM-ASR} (DiariZen) (EN 1600h)& 0.6B & 27.87 & 17.34 & 18.28 & 14.83 & 25.05 & 26.60 & 24.71 & 16.86 & 17.00 \\
        {DM-ASR} (DiariZen) (CN+EN 2900h) & 0.6B & 27.87 & 17.47 & 18.41 & 14.83 & 23.47 & 25.10 & 24.71 & 17.22 & 17.35 \\
        {DM-ASR} (DiariZen) (CN+EN 2900h) & 1.7B & 27.87 & 17.16 & 18.09 & 14.83 & 23.45 & 24.97 & 24.71 & 16.76 & 16.89 \\
        {DM-ASR} (S2SND) (CN+EN 2900h) & 1.7B & \cellcolor{gray!25}11.48 (4.89) & \cellcolor{gray!25}16.40 &\cellcolor{gray!25} 16.93 & \cellcolor{gray!25}13.72 (5.95) &\cellcolor{gray!25} 21.26 & 22.07&\cellcolor{gray!25}11.16 (6.53) & 15.91&\cellcolor{gray!25}16.10 \\
        {DM-ASR-L} (S2SND) (CN+EN 7500h) & 4.0B & 11.50 (4.90) & 17.56 & 18.16 & 13.74 (5.95) & 22.59 & 23.42 & 11.17 (6.52) & 16.83 & 17.00 \\
        \quad + GSPO (S2SND) (CN+EN 7500h) & 4.0B & 11.52 (4.91) & 16.72 & 17.14 & 13.73 (5.95) & 21.43 & \cellcolor{gray!25}22.02 & 11.18 (6.52) & 16.02 & 16.13 \\
        \bottomrule
	\end{tabular}
    }
	\end{threeparttable}
\end{table*}

\begin{table*}[t]
\centering\small
\setlength{\tabcolsep}{2pt}
\renewcommand{\arraystretch}{1.05}
\caption{Online results (\%). Mandarin columns report cpCER/tcpCER and English columns cpWER/tcpWER. DER uses a 0-s collar; tcpWER/tcpCER use a 5-s collar.}
\label{exp:online}
\resizebox{\textwidth}{!}{%
\begin{tabular}{l ccc ccc ccc ccc ccc}
\toprule
\multirow{2}{*}{\textbf{System / online setting}}
& \multicolumn{3}{c}{\textbf{AliMeeting}} & \multicolumn{3}{c}{\textbf{AISHELL-4}}
& \multicolumn{3}{c}{\textbf{AMI-IHM}} & \multicolumn{3}{c}{\textbf{AMI-SDM}}
& \multicolumn{3}{c}{\textbf{Fisher}} \\
\cmidrule(lr){2-4}\cmidrule(lr){5-7}\cmidrule(lr){8-10}\cmidrule(lr){11-13}\cmidrule(lr){14-16}
& DER & cpCER & tcpCER & DER & cpCER & tcpCER & DER & cpWER & tcpWER & DER & cpWER & tcpWER & DER & cpWER & tcpWER \\
\midrule
VibeVoice-ASR-Streaming (7B)~\cite{tu2026vibevoiceasrstreaming} & -- & 39.80 & -- & -- & 28.70 & -- & -- & 27.48 & -- & -- & 39.01 & -- & -- & -- & -- \\
JEDIS-LLM (VAD $\leq$3\,s, SPC $\leq$3\,s)~\cite{shi2025train} & -- & -- & -- & -- & -- & -- & -- & -- & -- & -- & -- & -- & -- & 21.03 & -- \\
DM-ASR-L & 10.64 & 18.09 & 18.40 & 10.61 & 17.01 & 17.53 & 12.07 & 18.09 & 18.65 & 15.00 & 23.96 & 24.79 & 15.22 & 19.36 & 19.62 \\
\quad+ GSPO & 10.67 & \cellcolor{gray!15}\textbf{17.53} & \cellcolor{gray!15}\textbf{17.76} & 10.67 & \cellcolor{gray!15}\textbf{15.43} & \cellcolor{gray!15}\textbf{15.78} & 12.12 & \cellcolor{gray!15}\textbf{17.38} & \cellcolor{gray!15}\textbf{17.83} & 15.05 & \cellcolor{gray!15}\textbf{22.86} & \cellcolor{gray!15}\textbf{23.52} & 15.25 & \cellcolor{gray!15}\textbf{18.76} & \cellcolor{gray!15}\textbf{18.94} \\

\bottomrule
\end{tabular}
}
\end{table*}

\section{Experiment Setup}
\subsection{Datasets}
\label{sec:datasets}
DM-ASR is trained on conversational speech from English and Mandarin corpora. The English portion includes the IHM-Mix and SDM-Mix versions of AMI~\cite{ami}, the SDM-Mix version of ICSI~\cite{icsi}, the English subsets of MLC-SLM~\cite{mlc}, and Fisher~\cite{cieri2004fisher}. The Mandarin portion includes AISHELL-4~\cite{aishell4}, AliMeeting~\cite{alimeeting}, MISP2025~\cite{misp2025}, HKUST~\cite{hkust}, MagicData-RAMC~\cite{ramc}, and conversational speech from Nexdata Technology Inc.\footnote{\url{https://www.nexdata.ai/datasets/speechrecog/1203}}. We do not use simulated training data.

We use several training mixtures: CN 212h contains AliMeeting and AISHELL-4; CN 630h adds MISP2025, MagicData-RAMC, and HKUST; CN 1300h further adds Nexdata. EN 630h contains AMI, ICSI, and one quarter of Fisher; EN 1600h adds another quarter of Fisher and the English MLC-SLM subset. CN+EN 2900h combines CN 1300h and EN 1600h. The final DM-ASR-L mixture, labeled CN+EN 7500h, additionally includes internal 2-speaker Chinese conversational data and the full fisher corpora.

We evaluate on AliMeeting Test and AISHELL-4 Test for Mandarin, and on the AMI-IHM, AMI-SDM, and Fisher test sets for English. Appendix~\ref{app:dataset_details}, Table~\ref{tab:dataset_composition}, summarizes source durations and training/evaluation roles; benchmark results are reported in Section~\ref{sec:offline_results}.

\subsection{Models and Training}
The compact models use Whisper-large-v3-turbo\footnote{\url{https://huggingface.co/openai/whisper-large-v3-turbo}} as the speech encoder, a trainable two-layer GELU MLP projector, with Gemma3-270M\footnote{\url{https://huggingface.co/google/gemma-3-270m-it}}, Qwen3-0.6B\footnote{\url{https://huggingface.co/Qwen/Qwen3-0.6B}}, or Qwen3-1.7B\footnote{\url{https://huggingface.co/Qwen/Qwen3-1.7B}} as decoders. Unless otherwise specified, pretrained backbones are adapted with LoRA ($r=16$, $\alpha=32$), while the projector is optimized jointly. Word-level timestamps are obtained using MFA~\cite{mfa}. Training clips span 15--25 seconds; the stated baseline recipe uses eight NVIDIA A6000 48GB GPUs, batch size 2 per GPU, AdamW, a $10^{-4}$ peak learning rate, and linear warmup-decay.

DM-ASR-L uses the Qwen3-ASR-1.7B\footnote{\url{https://huggingface.co/Qwen/Qwen3-ASR-1.7B}} encoder and a Qwen3-4B\footnote{\url{https://huggingface.co/Qwen/Qwen3-4B}} decoder; both backbones are frozen and adapted with LoRA ($r=16$, $\alpha=32$), and the projector is trained jointly with a $10^{-4}$ peak learning rate. Starting from DM-ASR-L, GSPO runs for 500 steps, with constant learning rates of $10^{-6}$; we denote the resulting model DM-ASR-L + GSPO. The Gemini baselines use Gemini-2.5-Pro\footnote{\url{https://ai.google.dev/gemini-api/docs/models?hl=zh-cn\#gemini-2.5-pro}} and Gemini-3.0-Pro\footnote{\url{https://ai.google.dev/gemini-api/docs/models?hl=zh-cn\#gemini-3-pro}}.

\subsection{Evaluation Metrics}
We use the MeetEval toolkit\footnote{\url{https://github.com/fgnt/meeteval}} to report diarization error rate (DER), concatenated minimum-permutation WER/CER (cpWER/cpCER), and time-constrained cpWER/CER (tcpWER/tcpCER). DER measures speaker attribution, cpWER/cpCER measure speaker-attributed lexical accuracy under speaker permutation, and tcpWER/tcpCER additionally enforce temporal matching. Lower scores are better. Unless otherwise stated, DER uses a 0-s collar and tcpWER/tcpCER use a 5-s collar. All scores are percentages. Detailed metric definitions and output-field evaluation settings are provided in Appendix~\ref{app:evaluation_protocols}.
All DM-ASR results are computed on complete recordings.

\subsection{Inference}
Offline results use the specified offline DiariZen or S2SND; online DM-ASR-L uses a separate online S2SND front end. All DM-ASR models and DM-ASR-L use greedy decoding.

\begin{table*}[t]
\centering\small
\setlength{\tabcolsep}{4pt}
\caption{Online recognition error rates and latency.}
\label{tab:latency_comparison}
\resizebox{\textwidth}{!}{%
\begin{tabular}{@{}llll@{}}
\toprule
Dataset & Systems & cpCER/cpWER (\%) & Avg. Latency \\
\midrule
AliMeeting & VibeVoice-ASR-Streaming / DM-ASR-L + GSPO & 39.80 / 17.53 & 2\,s / (0.68+1.5)\,s \\
Fisher & JEDIS-LLM / DM-ASR-L + GSPO & 21.03 (18.18) / 18.76 & 1.5\,s (2.5\,s) / (0.68+1.5)\,s \\
\bottomrule
\end{tabular}%
}
\end{table*}

\begin{table*}[t]
\centering
\setlength{\tabcolsep}{4pt}
\renewcommand{\arraystretch}{0.8}
\caption{Ablation on AliMeeting and AISHELL-4. Metrics are cpCER (\%) and tcpCER (\%).}
\label{exp:ali}
\resizebox{\textwidth}{!}{%
\begin{tabular}{llclcrrrr}
\toprule
\multirow{3}{*}{\textbf{ID}}
& \multirow{3}{*}{\textbf{Method}}
& \multirow{3}{*}{\textbf{\makecell[l]{Word-level \\Timestamps}}}
& \multirow{3}{*}{\textbf{\makecell[l]{Training \\Data}}}
& \multirow{3}{*}{\textbf{\makecell[l]{Max \\Duration(s)}}}
& \multicolumn{2}{c}{\textbf{AliMeeting}} & \multicolumn{2}{c}{\textbf{AISHELL-4}} \\
\cmidrule(l{2pt}r{2pt}){6-7} \cmidrule(l{2pt}r{2pt}){8-9}
&&&&&\textbf{cpCER} & \textbf{tcpCER} & \textbf{cpCER} & \textbf{tcpCER} \\
\midrule
M1 & DM-ASR (Gemma3 270m) & \xmark & CN 212h & 15 & 31.07 & 31.80 & 29.18 & 29.92 \\
M2 & DM-ASR (Gemma3 270m) & \cmark & CN 212h & 15 & 28.24 & 28.96 & 27.75 & 28.51 \\
M3 & DM-ASR (Gemma3 270m) & \cmark & CN 630h & 15 & 26.44 & 27.11 & 23.55 & 25.88 \\
M4 & DM-ASR (Gemma3 270m) & \cmark & CN 630h & 25 & 24.33 & 24.98 & 22.03 & 23.74 \\
M5 & DM-ASR (Qwen3 0.6B) & \cmark & CN 630h & 25 & 23.46 & 24.09 & 21.58 & 23.04 \\
M6 & DM-ASR (Qwen3 0.6B) & \cmark & CN 1300h & 25 & \cellcolor{gray!25}21.60 & \cellcolor{gray!25}22.23 & \cellcolor{gray!25}19.69 & \cellcolor{gray!25}20.40 \\
\bottomrule
\end{tabular}%
}
\end{table*}

\section{Experiment Results}
\subsection{Offline Recognition Performance}\label{sec:offline_results}
Tables~\ref{exp:man} and~\ref{exp:eng} compare DM-ASR with cascaded and unified systems on Mandarin and English benchmarks. Three patterns emerge: diarization quality affects speaker-attribution scores; the strongest Mandarin results come from a data-richer configuration with a larger decoder; and encoder suitability appears language-dependent.

\paragraph{Mandarin benchmarks.}
DM-ASR-L achieves cpCER/tcpCER of 17.51\%/17.78\% on AliMeeting and 15.58\%/15.95\% on AISHELL-4, outperforming the 1.7B S2SND variant in cpCER (21.40\% and 17.66\%, respectively). These improvements are obtained with a Qwen3-ASR-1.7B encoder, a 4B LLM decoder, and an expanded 7,500-hour training set. Among competing systems, MOSS Transcribe Diarize employs a compact 0.9B model trained on extensive in-the-wild data, although its total training hours are not reported~\cite{yu2026moss}. Qwen-Audio-3.1-ASR-Next reports cpCERs of 12.04\% and 13.51\% on AliMeeting and AISHELL-4, with 0.5-s-collar DERs of 2.15\% and 2.70\%, respectively~\cite{qwen_audio_3_1_asr_report}. Its audio-language pretraining uses approximately 20 million hours of data, and the duration of its speaker-diarized ASR training data is not reported. Moreover, Qwen-Audio-3.1-ASR-Next is evaluated on approximately 4-minute chunks, whereas DM-ASR is evaluated on complete recordings with speaker labels tracked throughout each recording; the results are therefore not directly comparable. Overall, DM-ASR-L demonstrates competitive Mandarin recognition performance with smaller data.

\paragraph{English benchmarks.}
Table~\ref{exp:eng} presents the results on AMI-IHM, AMI-SDM, and Fisher. Notably, although DiariZen exhibits a relatively high DER on AMI-IHM (27.87\%), DM-ASR still achieves a cpWER of 17.16\% with the 1.7B decoder, suggesting that the proposed framework can tolerate imperfect diarization cues without a proportional degradation in transcription accuracy. Replacing DiariZen with the stronger S2SND front end further reduces DER to 11.48\% and improves cpWER to 16.40\%, highlighting the benefits of more accurate diarization. These results suggest that DM-ASR can effectively exploit imperfect speaker-time cues while still benefiting from improvements in front-end diarization quality.
And we can also find that the Whisper-encoder configuration obtains cpWERs lower than DM-ASR-L. This pattern suggests that the Qwen3-ASR-1.7B encoder may be less suitable for English in our setting, while the Whisper encoder may be a better choice for English speech fine-tuning. DM-ASR-L nevertheless has lower cpWER than VibeVoice-ASR on the two AMI conditions.

\paragraph{Effect of GSPO.}
GSPO consistently improves DM-ASR-L on all five benchmarks, reducing cpCER/cpWER by 0.74--1.16 points and tcpCER/tcpWER by 0.78--1.40 points, with the largest gain on AMI-SDM (Tables~\ref{exp:man} and~\ref{exp:eng}). On English, it also narrows the gap to the Whisper-based 1.7B model to at most 0.32 cpWER points, partly compensating for the encoder mismatch noted above.

\subsection{Online Recognition Performance}
Table~\ref{exp:online} reports DM-ASR-L with an online S2SND front end and 25\,s endpoint-causal acoustic windows. It achieves cpCER/tcpCER of 18.09\%/18.40\% on AliMeeting and 17.01\%/17.53\% on AISHELL-4, and cpWER/tcpWER of 18.09\%/18.65\%, 23.96\%/24.79\%, and 19.36\%/19.62\% on AMI-IHM, AMI-SDM, and Fisher. Compared with the reported VibeVoice-ASR-Streaming results, DM-ASR-L achieves better results. The GSPO-tuned model further reduces cpCER to 17.53\% and 15.43\% on AliMeeting and AISHELL-4 and cpWER to 17.38\%, 22.86\%, and 18.76\% on AMI-IHM, AMI-SDM, and Fisher. These results demonstrate incremental \emph{segment-level} transcription.

\paragraph{Latency comparison.} Table~\ref{tab:latency_comparison} compares the recognition accuracy and latency of online systems. VibeVoice-ASR-Streaming has an estimated algorithmic delay of approximately 2.0\,s, based on its 2.9-s chunks and 0.5-s lookahead~\cite{tu2026vibevoiceasrstreaming}. For DM-ASR-L, the average segment-level ASR latency is 1.5\,s, obtained by averaging the latency across all finalized segments. Combined with the average online diarization delay of 0.68\,s, this yields an estimated total latency of 2.18\,s. On Fisher, DM-ASR-L + GSPO achieves 18.76\% cpWER, compared with JEDIS-LLM's 21.03\% at 1.5\,s and 18.18\% at 2.5\,s~\cite{shi2025train}. Unlike chunk-based decoding, DM-ASR-L leverages up to 25\,s of causal acoustic context without requiring future audio, enabling context-aware segment-level transcription.

\subsection{Ablation Study}
Table~\ref{exp:ali} isolates changes to timestamp supervision, training mixture, chunk duration, and model configuration on AliMeeting and AISHELL-4.

Adding word-level timestamp targets (M1--M2) lowers cpCER/tcpCER on both AliMeeting (31.07\%/31.80\% to 28.24\%/28.96\%) and AISHELL-4 (29.18\%/29.92\% to 27.75\%/28.51\%). Results also improve when the maximum clip length increases from 15\,s to 25\,s (M3--M4), and when training expands from CN 212h to CN 630h or CN 1300h (M2--M3, M5--M6), though the data mixtures change in composition as well as size. M4--M5 changes both decoder family and parameter count, so it does not isolate a model-scaling effect. The ablation supports timestamp supervision, longer context, and broader training as useful factors.

\section{Conclusion}
We presented DM-ASR, a diarization-conditioned multi-turn generation framework for structured multi-speaker speech recognition. By leveraging explicit speaker-time cues as structural priors, DM-ASR decomposes multi-speaker transcription into a sequence of speaker-time-conditioned queries, enabling context-aware recognition while preserving the shared mixed-speech input. The framework further supports optional word-level timestamp prediction and unifies offline and causal segment-level online decoding through flexible acoustic context and dialogue history management.

Experiments on Mandarin and English benchmarks demonstrate competitive performance across different model configurations and inference settings, highlighting the benefits of explicit diarization guidance and fine-grained temporal supervision. Sequence-level GSPO fine-tuning further aligns training with the speaker-attributed evaluation metrics and improves both offline and online recognition. The results also reveal the complementary roles of diarization and language models in speaker attribution, temporal grounding, and lexical recognition. Moreover, the same trained model supports incremental segment-level transcription using causal acoustic context, demonstrating the applicability of the proposed formulation to online scenarios. Future work will focus on improving robustness to imperfect diarization, maintaining speaker consistency across longer conversations, and reducing online decoding latency.

\section*{Limitations}
DM-ASR depends on the coverage and accuracy of its diarization front end. Speech that is missed by the front end is not issued as a transcription target, and incorrect speaker labels or segment boundaries can still harm speaker attribution, lexical recognition, and temporal grounding.

Online decoding commits one finalized segment-level segment at a time; it is not frame-synchronous streaming. Although the model uses only causal acoustic context, output time depends on diarization finalization and recognition decoding.



\bibliography{sample-base}

\clearpage
\appendix
\section{Diarization-conditioned Generation Details}
\label{app:generation}
\subsection{Special Token Inventory}
The model serializes speaker identities, window-relative timestamps, and output-format controls as discrete tokens. Speaker indices are remapped locally in each window; their original labels are retained for final reconstruction.
\begin{table}[t]
\centering\scriptsize
\begin{tabular}{lp{4.6cm}r}
\toprule
Category & Tokens & \# \\
\midrule
Audio & \texttt{<|start\_of\_audio|>}, \texttt{<|audio|>}, \texttt{<|end\_of\_audio|>} & 3 \\
Speaker & \texttt{<|start\_of\_spk|>}, \texttt{<|end\_of\_spk|>}, \texttt{<|spk\_idx\_0|>}\allowbreak--\allowbreak\texttt{<|spk\_idx\_19|>} & 22 \\
Time & \texttt{<|start\_of\_time|>}, \texttt{<|end\_of\_time|>}, \texttt{<|time\_idx\_0|>}\allowbreak--\allowbreak\texttt{<|time\_idx\_250|>} (0.1\,s), \texttt{<|time\_fine\_1|>}\allowbreak--\allowbreak\texttt{<|time\_fine\_4|>} (0.02--0.08\,s) & 257 \\
Output mode & \texttt{<|with\_timestamp|>} & 1 \\
\midrule
Total & & 283 \\
\bottomrule
\end{tabular}
\caption{Added tokens. \texttt{<|audio|>} is replaced by projected speech features; time indices are relative to the start of the acoustic window.}
\label{tab:special_tokens}
\end{table}

\subsection{Prompt and Response Examples}
\begin{figure*}[t]
\centering
\includegraphics[width=\linewidth]{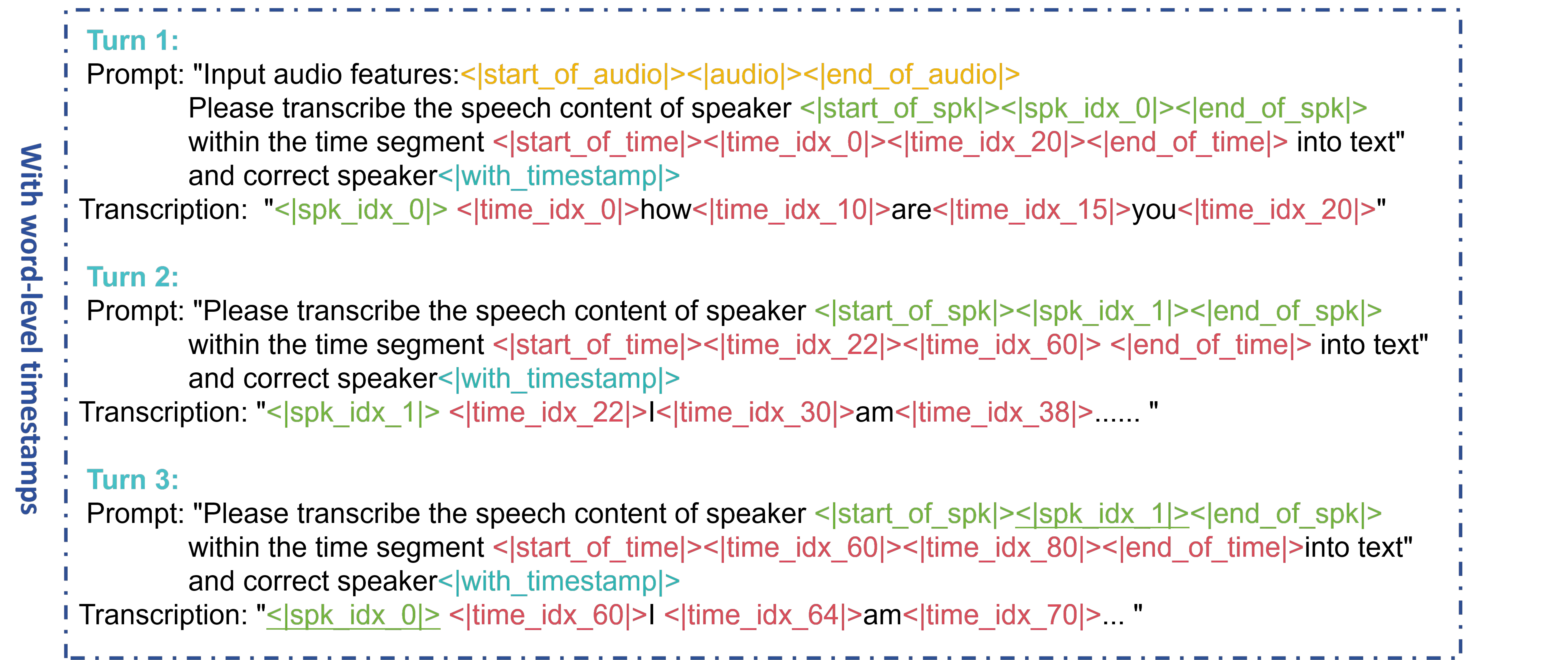}
\caption{Example of multi-turn diarization-conditioned prompts and responses. The first turn includes the audio representation, and subsequent turns add speaker--time queries using dialogue history. The optional timestamp mode produces word-level temporal tokens.}
\label{fig:prompt}
\end{figure*}

Figure~\ref{fig:prompt} illustrates how the prompt and response sequence is formed over three turns. The first turn contains the audio features and a speaker--time query; later turns add a new query without repeating the audio, while retaining the preceding dialogue context. The speaker indices are local to the acoustic window. With word-level timestamps enabled, the response includes timestamp tokens interleaved with the recognized words; the example also shows that a generated speaker token can differ from the speaker label in the query.

In Turn 3, the prompt may contain an incorrect speaker token due to perturbation, while the target transcription still begins with the correct speaker token. Through such training pairs, the model learns to revise the speaker identity in its output rather than mechanically copying the noisy prompt. The same idea also applies to the start and end timestamps: even when the temporal boundaries in the prompt are perturbed, the supervision remains anchored to the correct transcription target. This perturbation-based training can potentially improve the robustness of DM-ASR and enables it to better exploit context to correct imperfect diarization cues at inference time, especially when trained with more training data and stronger LLM backbones.

\section{Unified Offline and Online Decoding Details}
\label{app:online_decoding}
\subsection{Online segment-level Algorithm}
Each target enters the loop only after diarization finalizes it. The algorithm prunes history to previously committed turns whose start times fall within the current acoustic window, then commits and appends each generated response before processing subsequent targets. Speaker indices are local to each window and are mapped back to the original diarization labels for reconstruction. The speaker and timestamp tokens used in each prompt are listed in Appendix~\ref{app:generation}, Table~\ref{tab:special_tokens}.
\begin{algorithm}[t]
\caption{segment-level DM-ASR decoding}
\label{alg:online_decoding}
\begin{algorithmic}[1]
\Require Audio stream $X$, finalized diarization segments $D$, window length $L$
\State $H\gets\emptyset$
\For{each newly finalized target $(s_k,t_k^{\mathrm{st}},t_k^{\mathrm{ed}})$}
  \State $W_k\gets[\max(0,t_k^{\mathrm{ed}}-L),\,t_k^{\mathrm{ed}}]$
  \State Remove from $H$ the turns that start before $W_k$
  \State Build query $q_k$ with window-relative speaker--time tokens
  \State $r_k\gets \textsc{Generate}(X[W_k],q_k,H)$
  \State Commit $r_k$; append $(q_k,r_k)$ to $H$
\EndFor
\end{algorithmic}
\end{algorithm}

\subsection{History Management in Online Decoding}
\label{app:online_history}
Algorithm~\ref{alg:online_decoding} keeps a sliding dialogue history: when the causal window $W_k$ moves
forward, only the turns that start before $W_k$ are removed, and the remaining committed turns are kept as context.
A simpler alternative, denoted \emph{reset}, clears the whole history once the 25\,s window is full and restarts
the dialogue from the current target. Table~\ref{tab:online_history} compares the two strategies for the 1.7B
DM-ASR model (S2SND, CN+EN 2900h) under the same online S2SND front end.

The two strategies receive identical diarization inputs, and their DER differs by at most 0.04 points, so the
difference in cpCER comes from the lexical context retained in the dialogue history rather than from speaker
attribution. Sliding history lowers cpCER on all five test sets, with the largest gains on AliMeeting
($-$0.96) and Fisher ($-$0.78). Without it, each window boundary forces the decoder to restart without the preceding
conversation.

\begin{table*}[t]
\centering\small
\setlength{\tabcolsep}{5pt}
\caption{History management in online decoding for DM-ASR (1.7B, S2SND, CN+EN 2900h).
``Reset'' clears the dialogue history whenever the 25\,s window is full; ``sliding'' removes only the turns
that fall outside the current causal window (Algorithm~\ref{alg:online_decoding}).
Mandarin datasets use cpCER/tcpCER, while English datasets use cpWER/tcpWER (\%).}
\label{tab:online_history}
\begin{tabular}{l ccc ccc c}
\toprule
 & \multicolumn{3}{c}{\textbf{Reset}} & \multicolumn{3}{c}{\textbf{Sliding}} & \\
\cmidrule(lr){2-4}\cmidrule(lr){5-7}
\textbf{Dataset} & DER & \makecell{cpCER/\\cpWER} & \makecell{tcpCER/\\tcpWER} & DER & \makecell{cpCER/\\cpWER} & \makecell{tcpCER/\\tcpWER} & \makecell{$\Delta$cpCER/\\cpWER} \\
\midrule
AliMeeting & 10.67 & 22.32 & 22.69 & 10.69 & \textbf{21.36} & \textbf{21.73} & $-$0.96 \\
AISHELL-4  & 10.66 & 18.77 & 19.21 & 10.66 & \textbf{18.48} & \textbf{18.91} & $-$0.29 \\
AMI-IHM    & 12.09 & 17.08 & 17.62 & 12.12 & \textbf{16.68} & \textbf{17.21} & $-$0.40 \\
AMI-SDM    & 15.02 & 23.26 & 24.12 & 15.03 & \textbf{22.87} & \textbf{23.75} & $-$0.39 \\
Fisher     & 15.26 & 19.99 & 20.22 & 15.30 & \textbf{19.21} & \textbf{19.42} & $-$0.78 \\
\bottomrule
\end{tabular}
\end{table*}

\section{Dataset Details}
\label{app:dataset_details}
\subsection{Data Composition and Evaluation Splits}
\begin{table*}[t]
\centering\small
\caption{Composition and use of the training and evaluation corpora. Durations are source-level hours; speaker counts are per recording or session. Each training mixture uses only the portions specified in Section~\ref{sec:datasets}.}
\label{tab:dataset_composition}
\begin{tabularx}{\textwidth}{>{\raggedright\arraybackslash}X c r c X}
\toprule
Corpus & Language & Hours & Speakers & Use \\
\midrule
AMI (IHM-Mix, SDM-Mix) & English & 80 & 3--5 & Training; AMI-IHM/AMI-SDM evaluation \\
ICSI (SDM-Mix) & English & 66 & 3--10 & Training \\
MLC-SLM (English subsets) & English & 500 & 2 & Training \\
Fisher & English & 1,920 & 2 & Training; Fisher evaluation \\
AISHELL-4 & Mandarin & 107 & 3--7 & Training; AISHELL-4 evaluation \\
AliMeeting & Mandarin & 105 & 2--4 & Training; AliMeeting evaluation \\
MISP2025 & Mandarin & 119 & 4--8 & Training \\
HKUST & Mandarin & 149 & 2 & Training \\
MagicData-RAMC & Mandarin & 150 & 2 & Training \\
Nexdata & Mandarin & 672 & 3--10 & Training \\
Internal conversational data & Mandarin & 3,500 & 2 & Training \\
\bottomrule
\end{tabularx}
\end{table*}
Hours are approximate source durations reported in the main text; the named CN, EN, and bilingual training mixtures use the subsets described there. No simulated training data are used. The internal two-speaker corpus is included in DM-ASR-L's approximately 7,500-hour mixture.
\section{Evaluation Protocols}
\label{app:evaluation_protocols}
\subsection{Metric Definitions}
We use the MeetEval toolkit to score multi-speaker transcription. All reported values are percentages, and lower scores are better.

\textbf{Diarization Error Rate (DER)} measures speaker-attribution errors by accounting for missed speech, false alarms, and speaker confusion. Unless stated otherwise, our DER uses a 0-s forgiveness collar. Some cited baselines use a 0.5-s collar; the table captions identify these cases, and parenthesized values provide 0.5-s-collar results for comparison.

\textbf{Concatenated minimum-permutation WER/CER (cpWER/cpCER)} measures lexical accuracy while accounting for speaker attribution. Hypothesis and reference utterances are concatenated separately for each speaker, and the minimum word or character error rate is computed over speaker-label permutations. Thus, cpWER/cpCER evaluates what was recognized and the consistency of assigning it to speakers, but does not directly score timestamp accuracy. WER is used for English and CER for Mandarin.

\textbf{Time-constrained minimum-permutation WER/CER (tcpWER/tcpCER)} extends cpWER/cpCER by restricting word or character matching to a 5-s temporal collar. It therefore also penalizes temporal misalignment and reflects who spoke, what was said, and when it occurred. 

\subsection{Speaker and Timestamp Output Settings}
\label{app:cue_analysis}
We evaluate four combinations of the source used for the final speaker labels and timestamps:
\begin{itemize}
\item \textbf{Diarization speakers and diarization times:} both fields are taken from the diarization front end.
\item \textbf{Diarization speakers and model-predicted times:} speaker labels are retained from diarization, while the model predicts timestamps.
\item \textbf{Model-predicted speakers and diarization times:} timestamps are retained from diarization, while the model predicts speaker labels.
\item \textbf{Model-predicted speakers and model-predicted times:} the model predicts both final fields.
\end{itemize}
In all four settings, the input query remains diarization-conditioned; only the source of the final output fields changes. This comparison therefore measures output-field selection rather than removing diarization from the model input.

\begin{figure}[t]
\centering
  \IfFileExists{aishell4_der_cpcer_tcpcer_final.png}{\includegraphics[width=\linewidth]{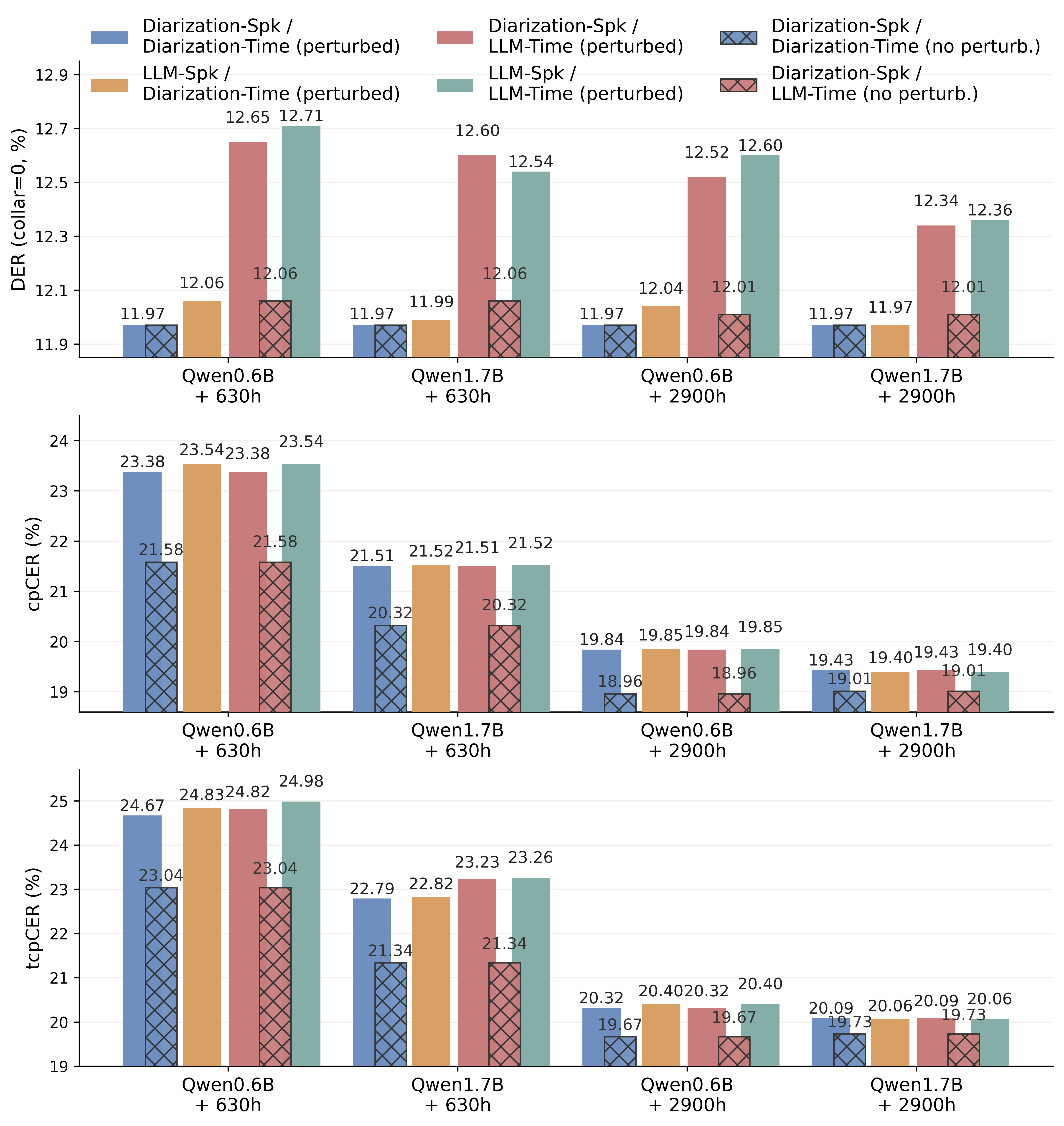}}{\fbox{AISHELL-4 speaker/time cue figure}}
  \caption{Performance across output-field settings on AISHELL-4, measured by DER, cpCER, and tcpCER, with and without diarization-label perturbation ($p=0.1$).}
  \label{fig:dia}
\end{figure}

\begin{table}[t]
\centering\small
\caption{Out-of-domain results on AISHELL-5.}
\label{tab:aishell5_ood}
\begin{tabularx}{\columnwidth}{>{\raggedright\arraybackslash}Xr}
\toprule
System  & cpCER (\%) \\
\midrule
SpeakerLM (7B, 7,639 h)~\cite{yin2025speakerlm} & 47.81 \\
DM-ASR (1.7B, CN+EN 2900h) & 39.79 \\
DM-ASR-L (4B, CN+EN 7500h) & 35.13 \\
\bottomrule
\end{tabularx}
\end{table}

\section{Additional Evaluation Results}
\label{app:additional_results}
\subsection{AISHELL-4 Cue Analysis}
Figure~\ref{fig:dia} compares different evaluation setups on AISHELL-4 in terms of DER, cpCER, and tcpCER under both perturbed and non-perturbed training conditions. Several clear trends can be observed. First, using diarization-provided speaker and time information consistently yields the best performance, especially on DER, showing that reliable external diarization cues remain highly beneficial for speaker attribution. Second, the degradation caused by replacing diarization
cues with LLM predictions is relatively limited, which suggests that the model can exploit contextual information to partially correct imperfect speaker and temporal information rather than simply depending on the prompt. Third, increasing the training data scale leads to substantial and consistent improvements across DER, cpCER, and tcpCER, in many cases exceeding the gains obtained by scaling model size alone. Fourth, the gap between different evaluation setups gradually narrows as both model size and training data scale increase. Finally, compared with the Qwen0.6B+630h setting, larger and better-trained models show more stable performance across all setups, particularly when both speaker labels and timestamps are predicted by the LLM, indicating stronger robustness and better cue refinement ability. This trend is also reflected in higher-resource settings (1.7B+2900h), where the benefit of perturbed training lies mainly in improving robustness under imperfect diarization cues, rather than further improving the best-case setup.

\subsection{Out-of-Domain Evaluation on AISHELL-5}
AISHELL-5~\cite{aishell5_2025} provides an out-of-domain evaluation set. Table~\ref{tab:aishell5_ood} reports the available SpeakerLM reference and DM-ASR cpCER.

\begin{table*}[t]
\centering\small
\setlength{\tabcolsep}{3.5pt}
\caption{Design-choice ablations for DM-ASR-L (A1, offline, Qwen3-4B decoder, 25\,s window). ``New rows only'' updates
only the embedding rows of the added special tokens; ``full embedding'' updates the whole embedding matrix.
A1 reports the DM-ASR-L results from Table~\ref{exp:man}.
Metrics are DER, cpCER, and tcpCER (\%).}
\label{tab:design_ablation}
\begin{tabular}{l l l l l ccc ccc}
\toprule
 & & & & & \multicolumn{3}{c}{\textbf{AliMeeting}} & \multicolumn{3}{c}{\textbf{AISHELL-4}} \\
\cmidrule(lr){6-8}\cmidrule(lr){9-11}
\textbf{ID} & \textbf{Encoder} & \textbf{New-token emb.} & \textbf{Data} & \textbf{Decoding}
 & DER & cpCER & tcpCER & DER & cpCER & tcpCER \\
\midrule
\multicolumn{11}{l}{\textit{A. Decoding strategy}} \\
\midrule
A0 & Qwen3-ASR & new rows only & CN+EN & non-overlapping & 10.06 & 17.90 & 18.18 & 9.70 & 15.87 & 16.36 \\
A1 & Qwen3-ASR & new rows only & CN+EN & sliding & 10.08 & \textbf{17.51} & \textbf{17.78} & 9.70 & \textbf{15.58} & \textbf{15.95} \\
\midrule
\multicolumn{11}{l}{\textit{B. Design-factor ablations}} \\
\midrule
B1 & Qwen3-ASR & full embedding & CN+EN & non-overlapping & 10.07 & 18.69 & 19.00 & 9.70 & 17.40 & 18.18 \\
B2 & Whisper-large-v3-turbo & new rows only & CN+EN & non-overlapping & 10.06 & 20.65 & 21.01 & 9.70 & 18.62 & 19.13 \\
B3 & Qwen3-ASR & new rows only & CN only & non-overlapping & 10.07 & 17.86 & 18.15 & 9.70 & 15.91 & 16.34 \\
\bottomrule
\end{tabular}
\end{table*}

\subsection{Design-choice Ablations for DM-ASR-L}
\label{app:design_ablation}
Table~\ref{tab:design_ablation} examines design choices of DM-ASR-L on AliMeeting and AISHELL-4. All rows use the Qwen3-4B decoder, a 25\,s window, and the same offline S2SND diarization input.
A0 and A1 compare decoding strategies; B1--B3 each change one design factor relative to A0.

\paragraph{Offline decoding.} A0 decodes consecutive non-overlapping 25\,s windows independently. A1 decodes densely
sliding windows, each conditioned on transcripts already predicted in earlier windows. The two give similar results
(17.90 vs.\ 17.51 cpCER on AliMeeting; 15.87 vs.\ 15.58 on AISHELL-4). Because each offline window already contains
a full 25\,s of context, carrying history across windows has a limited effect offline. By contrast,
history matters in online decoding (Appendix~\ref{app:online_history}).

\paragraph{New-token embeddings.} Updating only the embedding rows of the added special tokens (A0) outperforms
updating the full embedding matrix (B1) by 0.8 and 1.5 cpCER points on AliMeeting and AISHELL-4, suggesting that
keeping the pretrained vocabulary embeddings fixed preserves the decoder's language-modeling ability.

\paragraph{Speech encoder.} Replacing Qwen3-ASR-1.7B with Whisper-large-v3-turbo (B2) increases cpCER by 2.75 and
2.75 points. Together with the English results in Table~\ref{exp:eng}, this supports the observation in
Section~\ref{sec:offline_results} that encoder suitability is language-dependent.

\paragraph{Bilingual training.} Training on the Mandarin portion alone (B3) yields nearly the same Mandarin results
as the bilingual mixture (A0), so adding English data does not degrade Mandarin recognition.

\end{document}